\let\ifarxiv=\iftrue     
\pdfoutput=1

\ifarxiv

\documentclass[12pt,a4paper]{article}
\usepackage[a4paper,text={450pt,650pt},centering]{geometry}

\fi

\ifarxiv\else

\documentclass[11pt,a4paper]{article}
\usepackage{mathptmx}
\usepackage[a4paper,text={130mm,198mm}]{geometry}

\fi

\usepackage{amsmath,amssymb}
\usepackage[bookmarks=true,hyperfigures=true]{hyperref}
\usepackage{graphicx}
\usepackage[nosort]{cite}
\ifarxiv\usepackage[bulletsep]{collref}\fi

\let\oldbfseries=\bfseries
\let\oldmdseries=\mdseries
\let\oldnormalfont=\normalfont
\renewcommand{\bfseries}{\oldbfseries\boldmath}
\renewcommand{\mdseries}{\oldmdseries\unboldmath}
\renewcommand{\normalfont}{\oldnormalfont\unboldmath}

\allowdisplaybreaks[3]

\numberwithin{equation}{section}

\usepackage[font=small,labelfont=bf,width=0.85\textwidth]{caption}

\providecommand{\hypersetup}[1]{}

\hypersetup{plainpages=false}
\hypersetup{pdfpagemode=UseNone}
\hypersetup{bookmarksnumbered=true}
\hypersetup{pdfstartview=FitH}
\hypersetup{colorlinks=false}
\hypersetup{citebordercolor={.5 1 .5}}
\hypersetup{urlbordercolor={.5 1 1}}
\hypersetup{linkbordercolor={1 .7 .7}}

\providecommand{\href}[2]{#2}
\providecommand{\arxivlink}[1]{\href{http://arxiv.org/abs/#1}{arxiv:#1}}

\begin{document}


\thispagestyle{empty}
\phantomsection
\addcontentsline{toc}{section}{Title}

\begin{flushright}\footnotesize%
\texttt{\arxivlink{1012.3997}}\\
overview article: \texttt{\arxivlink{1012.3982}}%
\vspace{1em}%
\end{flushright}

\begingroup\parindent0pt
\begingroup\bfseries\ifarxiv\Large\else\LARGE\fi
\hypersetup{pdftitle={Review of AdS/CFT Integrability, Chapter IV.1: Aspects of Non-Planarity}}%
Review of AdS/CFT Integrability, Chapter IV.1:\\
Aspects of Non-Planarity
\par\endgroup
\vspace{1.5em}
\begingroup\ifarxiv\scshape\else\large\fi%
\hypersetup{pdfauthor={C. Kristjansen}}%
C.\ Kristjansen
\par\endgroup
\vspace{1em}
\begingroup\itshape
The Niels Bohr Institute, Blegdamsvej 17, DK-2100 Copenhagen \O,
Denmark
\par\endgroup
\vspace{1em}
\begingroup\ttfamily
kristjan@nbi.dk
\par\endgroup
\vspace{1.0em}
\endgroup

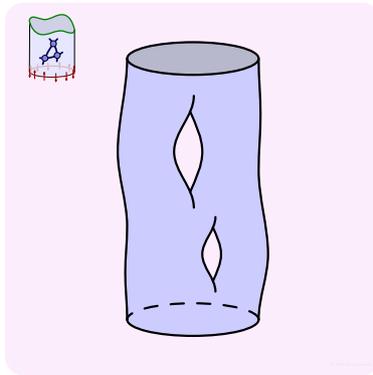
\begin{center}
\includegraphics[width=5cm]{TitleIV1.mps}
\vspace{1.0em}
\end{center}

\paragraph{Abstract:}
We review the role of integrability in certain
aspects of ${\cal N}=4$ SYM which go beyond the planar spectrum. 
In particular, we discuss integrability in relation to
non-planar anomalous dimensions, multi-point
functions and Maldacena-Wilson loops.

\ifarxiv\else
\fi
\hypersetup{pdfsubject={MSC (2010): 81T13, 81T30, 81T60, 81R12.}}%
\ifarxiv\else
\fi
\hypersetup{pdfkeywords={N=4 SYM, AdS/CFT, 
non-planar anomalous dimensions, three-point functions, 
Maldacena-Wilson loops, integrability.}}%

\newpage




\section{Introduction}

The discovery of the integrability of the planar spectral problem of 
AdS/CFT~
\cite{Minahan:2002ve,Beisert:2003tq,Mandal:2002fs,Bena:2003wd}
has provided us with a wealth of new results and tools for the study of
gauge and string theory. Given this success it is natural to investigate  
whether the integrability extends to other aspects of the AdS/CFT 
correspondence. Here we shall discuss this possibility mainly from 
the gauge theory perspective and staying entirely within the maximally
supersymmetric gauge theory in four dimensions, ${\cal N}=4$ SYM. 
The fate of the integrability of the planar spectral problem when reducing or
completely removing the supersymmetry is
discussed in the
chapters~\cite{chapDeform} and~\cite{chapQCD}. A natural direction in which
to search for integrability is in the non-planar version of the spectral
problem. As we will review below, while the non-planar version of the
dilatation generator can easily be written down (at least in some sub-sectors
and to a certain loop order) attempts to 
diagonalize it have so far not revealed 
any traces of integrability. For a conformal field theory like
${\cal N}=4$ SYM natural observables apart from anomalous dimensions 
are the structure constants which appear in the three point functions
of the theory and govern the theory's operator product expansion.
Three-point functions are of course not unrelated to 
non-planar anomalous dimensions as correlators of three traces
can be seen
as building blocks for higher genus two-point functions.
 As we shall see the calculation of
structure constants of ${\cal N}=4$ SYM is impeded by extensive
operator mixing. For a certain subset of operators, 
this mixing can be
handled via the diagonalization of the planar dilatation operator
and the structure constants can be calculated using tools pertaining to planar
integrability. An integrable structure allowing to treat all types
of three-point functions has not been identified.

%

Anomalous dimensions and structure constants are observables which are
associated with local gauge invariant operators but in a gauge theory 
one of course also has at hand numerous types of non-local 
observables such as Wilson loops, 't Hooft loops, surface operators and
domain walls. 
Here we will limit our discussion to Wilson loops, more precisely to
locally supersymmetric Maldacena-Wilson loops. Another type of
Wilson loops, Alday-Maldacena-Wilson loops and their relation to scattering
amplitudes of ${\cal N}=4$ SYM will be discussed 
in the chapters~\cite{chapAmp,chapDual,chapTDual}. 
As was known before the discovery of the spin-chain related integrability
of the AdS/CFT system, expectation values of 
Maldacena-Wilson loops can in certain cases be expressed in terms of 
expectation values of a
zero-dimensional integrable matrix model and this connection has provided us
with the most successful test of the AdS/CFT correspondence 
beyond the planar limit to date.
The connection of Maldacena-Wilson
loops to integrability in the form of spin-chain integrability is so far
very limited.

We start by discussing the role of integrability in connection
with non-planar anomalous dimensions in section~\ref{nonplanar} and
subsequently treat multi-point functions and Maldacena-Wilson loops
in sections~\ref{multipoint} and~\ref{Wilsonloops}.

\section{Non-planar anomalous dimensions \label{nonplanar}}

In a CFT conformal operators, $\{{\cal O}_{\alpha}\}$, and their associated 
conformal dimensions, $\Delta_{\alpha}$, 
are characterized by being eigenstates and eigenvalues of the dilatation
generator, $\hat{D}$. As a consequence of this two-point functions of
conformal operators upon appropriate normalization take the form
\begin{equation}
\langle {\cal O}_{\alpha}(x) {\cal O}_{\beta}(y) \rangle =
\frac{\delta_{\alpha \beta}}{(x-y)^{2\Delta_{\alpha}}}.
\end{equation}

\subsection{The non-planar dilatation generator}
The dilatation generator, 
${\hat {D}}$, of ${\cal N}=4$ SYM has a double expansion 
in $\lambda$ and $\frac{1}{N}$ where $\lambda$ is the 't Hooft coupling
which we until further notice take to be
\begin{equation}
\lambda=\frac{g_{\mbox{\tiny YM}}^2N}{8\pi^2},
\end{equation}
and where $N$ is the order of the gauge group, $SU(N)$. By the planar 
limit we mean the limit $N\rightarrow \infty$, $\lambda$ fixed.
At a finite order in $\lambda$ the $\frac{1}{N}$-expansion of the
dilatation generator starts at order $N^0$ and
terminates after finitely many
terms, the number of which increases with the loop order. 
The planar dilatation generator and its loop expansion is discussed
in the chapter~\cite{chapHigher}. 
The non-planar part of the dilatation generator 
was first derived 
at one loop order in the $SO(6)$ sector~\cite{Beisert:2002bb,Beisert:2002ff},
see also~\cite{Constable:2002hw}. 
The derivation was based on evaluation of 
Feynman diagrams and was extended to two-loop
order in the $SU(2)$ sector in~\cite{Beisert:2003tq}. Later a derivation
based entirely on algebraic arguments gave the dilatation generator
including non-planar parts for all fields at one-loop 
order~\cite{Beisert:2003jj} and for 
the fields in the $SU(1,1|2)$ sector at two-loop order~\cite{Zwiebel:2005er}.
Recently, the non-planar part of the dilatation generator was 
written down
at order $\lambda^{3/2}$ in the $SU(2|3)$ sector~\cite{Xiao:2009pv}.
In addition, the non-planar part of the dilatation
generator is known in the scalar
sector in a certain ${\cal N}=2$ superconformal gauge 
theory~\cite{DeRisi:2004bc}.
In ABJM theory~\cite{Aharony:2008ug} and 
ABJ theory~\cite{Aharony:2008gk} the non-planar part of the
two-loop dilatation generator has been derived
in a $SU(2)\times SU(2)$ 
sector~\cite{Kristjansen:2008ib,Caputa:2009ug}.\footnote{We remark that
our $\hat{D}$ is the dilatation generator describing the asymptotic spectrum. 
Hence we ignore the wrapping contributions discussed in
the chapters~\cite{chapHigher,chapLuescher,chapTBA,chapTrans}. 
In particular, the splitting of the dilatation
operator into planar and non-planar parts that we discuss here pertains
to the asymptotic regime. What is here referred to as non-planar parts
of the dilatation generator might for short operators give rise to
planar wrapping contributions~\cite{Sieg:2005kd}.}

The diagonalization problem for the full dilatation generator of
${\cal N}= 4$ SYM has mainly been
studied in the $SU(2)$-sector which consists of multi-trace operators built
from two complex scalar fields, say $X$ and $Z$. For simplicity we shall 
likewise focus our discussion on this sector. The one-loop
dilatation generator including the non-planar parts 
 reads for the $SU(2)$ sector
\begin{equation}
\hat{D}= -\frac{\lambda}{N}\,
:\mbox{Tr}[X,Z][\check{X},\check{Z}]:, 
\hspace{0.7cm} \mbox{where} \hspace{0.7cm}
\check{Z}_{\alpha\,\beta} =
\frac{\delta}{\delta Z_{\beta\alpha}},
\end{equation}
and similarly for $\check{X}$. The normal ordering symbol signifies that
the derivatives should not act on the $X$ and $Z$ field belonging to
the dilatation generator itself.
Below we illustrate how the full dilatation generator acts on a double
trace operator. Notice that we only consider one out of four terms contributing
to the dilatation generator and that we only represent one possible way of 
applying the derivatives

\ifarxiv
\begin{picture}(320,50)(0,0)
\put(0,0){$\mbox{Tr}(ZX \check{Z}\check{X}) \cdot
\mbox{Tr}(X Z XX Z)\, \mbox{Tr}(X Z)=
\mbox{Tr}(Z X \check{Z} Z XX Z)\, \mbox{Tr}(XZ)$}
\put(51,20){\line(1,0){41}}
\put(51,20){\line(0,-1){7}}
\put(92,20){\line(0,-1){7}}
\put(242,17){\line(1,0){7}}
\put(242,17){\line(0,-1){4}}
\put(249,17){\line(0,-1){4}}
\put(252,13){\tiny 1}
\put(241,19){\line(1,0){38}}
\put(241,19){\line(0,-1){6}}
\put(279,19){\line(0,-1){6}}
\put(282,13){\tiny 2}
\put(239,21){\line(1,0){84}}
\put(239,21){\line(0,-1){8}}
\put(323,21){\line(0,-1){8}}
\put(326,13){\tiny 3}
\put(10,-30){$=N\mbox{Tr}(ZXXXZ)\,\mbox{Tr}(XZ)+ 
\mbox{Tr}(ZX)\, \mbox{Tr}(ZXX)\,\mbox{Tr}(XZ)+\mbox{Tr}(ZXZZZXXZ).$}
\end{picture}
\vspace*{1.5cm}

\fi

\ifarxiv\else

\begin{picture}(320,50)(0,0)
\put(0,0){$\mbox{Tr}(ZX \check{Z}\check{X}) \cdot
\mbox{Tr}(X Z XX Z)\, \mbox{Tr}(X Z)=
\mbox{Tr}(Z X \check{Z} Z XX Z)\, \mbox{Tr}(XZ)$}
\put(39,20){\line(1,0){32}}
\put(39,20){\line(0,-1){7}}
\put(71,20){\line(0,-1){7}}
\put(190,17){\line(1,0){6}}
\put(190,17){\line(0,-1){4}}
\put(196,17){\line(0,-1){4}}
\put(199,13){\tiny 1}
\put(188,19){\line(1,0){28}}
\put(188,19){\line(0,-1){6}}
\put(216,19){\line(0,-1){6}}
\put(219,13){\tiny 2}
\put(186,21){\line(1,0){64}}
\put(186,21){\line(0,-1){8}}
\put(250,21){\line(0,-1){8}}
\put(253,13){\tiny 3}
\put(10,-30){$=N\mbox{Tr}(ZXXXZ)\,\mbox{Tr}(XZ)+ 
\mbox{Tr}(ZX)\, \mbox{Tr}(ZXX)\,\mbox{Tr}(XZ)+\mbox{Tr}(ZXZZZXXZ).$}
\end{picture}
\vspace*{1.5cm}

\fi

\noindent
As is evident from this example the full one-loop dilatation generator can 
be written as follows
\begin{equation}
\hat{D}=\lambda(\hat{D}_0+\frac{1}{N}\,\,\hat{D}_++\frac{1}{N}\,\,\hat{D}_-),
\label{Doneloop}
\end{equation}
where $\hat{D}_+$ and $\hat{D}_-$ respectively increases and decreases
the trace number by one and where $\hat{D}_0$ conserves the number of traces.
Suggestions for how
to write $\hat{D}_+$ and $\hat{D}_-$ in a more explicit form can be found
in~\cite{Bellucci:2004ru,Peeters:2004pt}.
We notice that for gauge group $SO(N)$ or $Sp(N)$ the one-loop dilatation 
operator will have a term which is of order $\frac{1}{N}$ but still conserves
the number of traces~\cite{Caputa:2010ep}. At $l$-loop order the dilatation 
operator can change the number of traces by at most $l$. 
Notice that since the anomalous
dimensions are the {\it eigenvalues} of the dilatation generator these do not
necessarily have a $\frac{1}{N}$-expansion which truncates. What is more,
some anomalous dimensions do not even have a well-defined double expansion 
in $\lambda$ and $\frac{1}{N}$. An example of an operator with this property 
can be found in~\cite{Beisert:2003tq}. Speaking about a one-loop anomalous
dimension, however, always makes sense.
To calculate the leading $\frac{1}{N}$-corrections
to one-loop anomalous dimensions one can make use of standard quantum 
mechanical perturbation theory.
Let us assume that
we have found an eigenstate of the planar dilatation generator $\hat{D}_0$, 
i.e.
\begin{equation}
\hat{D}_0 |{\cal O}\rangle = \gamma_{\cal O} |{\cal O} \rangle,
\end{equation}
and let us treat the terms sub-leading in $\frac{1}{N}$ 
as a perturbation. First, let us assume that there are no degeneracies
between $n$-trace states and $(n+1)$-trace states
in the spectrum.
 If that is the case we can proceed by using non-degenerate
quantum mechanical perturbation theory. Clearly, the $\frac{1}{N}$
terms in eqn.~(\ref{Doneloop}) do not 
have any diagonal components so the correction to the
anomalous dimension
for the state $|{\cal O}\rangle$ reads
\begin{equation}
\delta \gamma_{\cal O}= \frac{1}{N^2}\,\sum_{{\cal K}\neq {\cal O}}
\frac{\langle {\cal O} | \hat{D}_+ + \hat{D}_-| {\cal K}\rangle
\cdot \langle {\cal K} | \hat{D}_+ + \hat{D}_-| {\cal O}\rangle
}{\gamma_{\cal O}-\gamma_{\cal K}},
\end{equation}
and is of order $\frac{1}{N^2}$.
If there are degeneracies between $n$-trace states and $(n+1)$-trace
states we have to diagonalize the
perturbation in the subset of degenerate states and the corrections
will typically be of order $\frac{1}{N}$. We remark that 
the dilatation generator is {\it not}
a Hermitian operator but it is related to its Hermitian conjugate 
by a similarity transformation and therefore its eigenvalues are always
real~\cite{Gross:2002mh,Janik:2002bd,Beisert:2002ff}.

\subsection{The non-planar spectrum and integrability \label{Dnonplanar}}

Planar ${\cal N}=4$ SYM is described in terms of only one parameter, 
$\lambda$,
and planar anomalous dimensions have a 
perturbative expansion in terms of this 
single parameter. This fact made it possible initially 
to search for integrability in
the planar spectrum order by order in $\lambda$. In particular, 
the concept of perturbative integrability was introduced,
meaning that at $l$ loops the planar spectrum could be described as an 
integrable system when 
disregarding terms of order $\lambda^{l+1}$~\cite{Beisert:2003tq}.
Studying this perturbative form of integrability eventually led to
the all loop Bethe equations conjectured to be true perturbatively
to any loop order and 
non-perturbatively as well~\cite{Beisert:2005fw,Beisert:2006ez,Beisert:2006ib}.
When going beyond the planar limit it is natural to follow
a similar perturbative approach. 
The question of integrability beyond the planar limit
has so far been addressed only perturbatively in $\frac{1}{N}$ 
at the one-loop order.
The fact that the non-planar part of the dilatation generator
introduces splitting and joining of
traces enormously enlarges the Hilbert space of states of the system.
This complicates the direct search for integrability via the identification
of conserved charges or the construction of an asymptotic 
S-matrix with the appropriate
properties. As a simple way of getting an indication of whether integrability
persists at the non-planar level one can test for degenerate 
parity pairs~\cite{Beisert:2003tq}. 
Parity pairs are operators with the same anomalous dimension but opposite
parity where the parity operation on a single trace operator is defined 
by~\cite{Doikou:1998jh}
\begin{equation}
\hat{P}\cdot \mbox{Tr}(X_{i_1}\, X_{i_2}\ldots X_{i_n})=
\mbox{Tr}(X_{i_n}\ldots X_{i_2} \, X_{i_1}).
\end{equation}
(For a multi-trace operator, $\hat{P}$ must act on each of its single trace
components.) At the planar one-loop level one observes a lot of such 
parity pairs. The presence of these degeneracies has its origin
in the integrability of the model. ${\cal N}=4$ SYM  is parity invariant
and its dilatation generator commutes with the parity operation, i.e.
\begin{equation}
[\hat{D},\hat{P}]=0.
\end{equation}
Notice that this only tells us that eigenstates of the dilatation generator
can be organized into eigenstates of the parity operator and nothing about
degeneracies in the spectrum. The degeneracies can be explained by the 
existence of an extra conserved charge, $\hat{Q}_3$, which commutes with the
dilatation generator but anti-commutes with parity, i.e.
\begin{equation}
[\hat{D},\hat{Q}_3]=0,\hspace{0.7cm} \{\hat{P},\hat{Q}_3\}=0.
\end{equation}
Acting on a state
 with $\hat{Q}_3$, one obtains another state with the opposite parity 
but with the same energy\footnote{There exist states which are unpaired
and annihilated by $\hat{Q}_3$.}. Taking  into account non-planar corrections
the degeneracies are lifted. Since parity is still conserved this is taken
as an indication (but not a  proof, obviously) of the disappearance of the
higher conserved charges and thus a breakdown of integrability. Notice
that in accordance with this picture,
the parity pairs survive the inclusion of planar higher loop corrections.
The situation in ABJM theory is the same. Degenerate parity pairs are 
seen at the planar level but disappear once non-planar corrections are 
taken into account~\cite{Kristjansen:2008ib}. 
(For ${\cal N}=4$ SYM with gauge group $SO(N)$ or
$Sp(N)$ parity is gauged and the concept of planar parity pairs loses
its meaning~\cite{Caputa:2010ep}. For ABJ theory parity is broken at the non-planar
level~\cite{Caputa:2009ug}.) Hence it seems that one can not hope for 
integrability of the spectrum of AdS/CFT 
beyond the planar limit, at least not 
in a simple
perturbative sense.\footnote{The paper, \cite{deMelloKoch:2009zm}, entitled 
``Hints of Integrability Beyond the Planar Limit:Non-trivial Backgrounds'' 
is dealing with 
anomalous dimensions of operators from the $SU(2)$-sector
consisting of the factor $(\det(Z))^M$  multiplying a single trace operator.
In the limit $N,M\rightarrow \infty$ with $\frac{N}{M}\rightarrow 0$
and $g_{\mbox{\tiny YM}}^2 M$ fixed the authors find a set of conserved
charges commuting with the dilatation generator. We remark, however, that
 in the limit
considered the terms $\hat{D}_+$ and $\hat{D}_-$ do not contribute to
the dilatation generator.}

\subsection{Results on non-planar anomalous dimensions}

Prior to the derivation of the dilatation generator of ${\cal N}=4$ SYM
anomalous dimensions were determined through a rather complicated
process which involved for each set of operators considered
an explicit calculation of their two-point correlation functions
through Feynman diagram evaluation. Early results on non-planar anomalous
dimensions for short operators 
obtained by this method can be found 
in~\cite{Penati:2001sv,Ryzhov:2001bp,Bianchi:2002rw,Arutyunov:2002rs}.

With the derivation of the dilatation generator the calculation of 
anomalous dimensions was enormously simplified. At the planar level 
one now even has at hand the tools of integrability and all information about
the (asymptotic) spectrum is encoded in a set of algebraic Bethe equations.
As argued above similar tools are not currently available at the 
non-planar level. Thus to obtain spectral information beyond
the planar limit one has to explicitly
diagonalize the dilatation generator in each closed subset of states.
For the following discussion it is convenient to
divide the set of operators into three different types, short
operators, BMN type operators and operators dual to spinning strings.

By short operators we mean operators which contain a finite, small number 
of fields. Such operators only mix with a finite, small number of other 
operators and the resulting mixing matrix can be calculated and diagonalized
by hand (or using Mathematica). Various results on non-planar corrections
to anomalous dimensions of short
operator in the $SU(2)$ sector of ${\cal N}=4$ SYM can be 
found in~\cite{Beisert:2003tq} and~\cite{Bellucci:2004ru}. 
Reference~\cite{Bellucci:2004ru} in addition
contains results on the $SL(2)$-sector
of ${\cal N}=4$ SYM. Results for the $SU(2)\times SU(2)$ sector
of ABJM and ABJ theory were obtained in~\cite{Kristjansen:2008ib} 
and~\cite{Caputa:2009ug}.

BMN type operators~\cite{Berenstein:2002jq} 
are operators consisting of many fields of 
one type and a few excitations
 in the form of fields of another type 
(or of derivatives). Two-excitation eigenstates can easily be written down
at the planar level. In the $SU(2)$ sector they read
\begin{equation}
{\cal O}_n^{J_0,J_1,\ldots,J_k}=\frac{1}{J_0+1}\sum_{p=0}^{J_0} \cos
\left(\frac{\pi n(2p+1)}{J_0+1} \right)\mbox{Tr}(X\,Z^p\, X\,Z^{J_0-p})
\mbox{Tr}(Z^{J_1})\ldots \mbox{Tr}(Z^{J_k}),
\end{equation}
where $0\leq n\leq \left[\frac{J_0}{2}\right]$
and where the corresponding planar eigenvalues are
\begin{equation}
E_n=8\lambda \sin^2(\frac{\pi n}{J_0+1}).
\end{equation}
Acting with the non-planar part of the dilatation generator on BMN states
only requires a finite and small number of operations and the non-planar
part of the mixing matrix for BMN states can easily be written 
down~\cite{Beisert:2002ff}.
Treating $\hat{D}_+ +\hat{D}_-$ as a perturbation of $\hat{D}_0$ one 
should thus be
able to determine the leading non-planar corrections to the anomalous
dimensions of BMN operators by standard quantum mechanical perturbation
theory, cf.\ section~\ref{Dnonplanar}. However, degeneracies between 
single and multiple-trace states require the use of degenerate perturbation
theory and due to the complexity of the coupling between degenerate states
the mixing problem for BMN states was never resolved. For a discussion of
this problem, see \cite{Freedman:2003bh,Kristjansen:2003uy,Gutjahr:2004qj}.
There is one case, however, for which there is no degeneracy issue and that
is for states with mode number, $n=1$. Here it is possible to find the 
leading non-planar correction to the anomalous dimension in the limit
$J_i\rightarrow \infty$, $i=0,1,\ldots,k$, and $\lambda \rightarrow \infty$
with $\lambda'=\lambda/J^2$ and $g_2=J^2/N$ fixed where
$J=\sum_{i=0}^k J_i$. 
The result reads~\cite{Beisert:2002bb,Constable:2002vq}
\begin{equation}
\delta E_{n=1}=\lambda' g_2^2\left(\frac{1}{12}+\frac{35}{32\pi^2}\right).
\label{BMNnonplanar}
\end{equation}
There exist similar results for BMN operators belonging to the $SL(2)$ sector
of ${\cal N}=4$ SYM~\cite{Gursoy:2002yy} and for BMN operators in  
a certain ${\cal N}=2$ superconformal
gauge theory~\cite{DeRisi:2004bc}. The result in eqn.~(\ref{BMNnonplanar})
was extended to two-loop order in~\cite{Beisert:2002ff}.

The third class of operators, operators dual to spinning strings, consist of
an infinitely large number of background fields and an infinite number of
excitations. In the $SU(2)$ sector they take the form
\begin{equation}
{\cal O}=\mbox{Tr}(Z^{J-M} X^M)+.\ldots,
\end{equation}
where $\ldots$ denotes similar terms obtained by permuting the fields and
where $J,M\rightarrow \infty$, but $M/J$ is kept finite. Acting with the
non-planar dilatation generator on such an operator involves an infinite
number of operations and becomes unfeasible. In~\cite{Casteill:2007td}, 
based on a  coherent state formalism, matrix elements of the non-planar 
dilatation generator between operators dual to particular folded spinning
strings were calculated but an explicit diagonalization of
the non-planar dilatation generator for the situation in question
did not seem tractable. 

\subsection{Comparison to string theory \label{comparison1}}

In order to generate string theory data with which to compare non-planar
corrections to anomalous dimensions one needs to take into account
string loop corrections corresponding to considering string world-sheets
of higher genus. For short operators such a comparison is currently
out of sight since we do not even have any examples of a successful
comparison at the planar level, except for certain BPS states which can
be shown to have vanishing anomalous 
dimensions~\cite{D'Hoker:1998tz}. Recently, it was shown at one-loop
order that certain 
1/4 BPS states can be labeled by irreducible representations of the
Brauer algebra~\cite{Kimura:2010tx}, see also~\cite{Brown:2010pb}.

 The situation is slightly more encouraging in the case of BMN operators.
Considering the BMN limit on the gauge theory side corresponds on the
string theory side to taking the Penrose limit of the $AdS_5 \times S^5$
background which turns the geometry into a PP-wave. On the PP-wave 
one can quantize the free IIB string theory in light cone gauge
and find the corresponding free spectrum. In addition, considering higher
genus effects is possible by means of light cone string field theory (LCSFT).
 A review of the PP-wave/BMN correspondence including
an introduction to LCSFT can be found in the 
references~\cite{Pankiewicz:2003pg,Plefka:2003nb,Spradlin:2003xc,Sadri:2003pr,
Russo:2004kr}.
In LCSFT string interactions are described in terms
of a three-string vertex which encodes the information about the splitting
and joining of strings. There seems to be several ways of consistently 
defining this three-string vertex and there exist at least three proposals
for its exact form. For all proposals, however, it holds that there is 
a freedom of choosing a certain pre-factor of the vertex. 
Reference~\cite{Grignani:2006en} constitutes the most recent review of
this topic describing the different possible choices of the three-vertex and 
containing all the relevant references. Furthermore, the authors 
of~\cite{Grignani:2006en}
show that
the one-loop gauge theory result~(\ref{BMNnonplanar}) 
can be obtained from LCSFT provided one chooses one particular of
the proposed vertices and chooses its pre-factor in a specific 
way.\footnote{It should be noticed, though, that the match to the 
one-loop gauge theory result is obtained after a truncation to the
so-called impurity conserving channel while at the same time
it is proved that generically all channels would contribute to the result.
In addition, it is pointed out that an undetermined supercharge could
potentially also contribute to the result.} 
It is, however, not possible to recover the two-loop gauge theory
result from the LCSFT and generically LCSFT
gives rise to half-integer powers of $\lambda'$
appearing in the expressions for non-planar anomalous dimensions.
Such half-integer powers of $\lambda'$ were also found in the analysis
of worldsheet one-loop corrections to the planar energies of spinning
strings~\cite{Beisert:2005cw} and eventually led to the 
recognition that the BMN
expansion breaks down not only at strong coupling but also at 
weak coupling starting at four-loop 
order~\cite{Eden:2006rx,Beisert:2006ez,Bern:2006ew}. Hence, it appears
that in order to obtain complete agreement between gauge and
string theory we are forced to consider
the full $AdS_5\times S^5$ geometry.

Finally, in the case of operators dual to spinning strings no direct 
comparison between gauge theory and string theory has been possible. 
In reference~\cite{Peeters:2004pt} the decay of a single folded spinning 
string into two such strings
was studied in a semi-classical approximation and
a certain relation between the conserved charges of the
decay products was found. If the semi-classical decay channel were the 
dominant one, as it is known to be in flat space, one could hope that 
the matrix elements for string splitting and joining found 
in~\cite{Casteill:2007td} could
encode some similar relation. The analysis of~\cite{Casteill:2007td}, however,
did not point towards the semi-classical decay channel being the
dominant one.

\section{Multi-point functions \label{multipoint}}
By multi-point functions we mean correlation functions of the following type
\begin{equation}
\langle {\cal O}_{\Delta_1}(x_1) {\cal O}_{\Delta_2}(x_2)\ldots 
{\cal O}_{\Delta_n}(x_n)\rangle,
\end{equation}
where the operators involved are eigenstates of the dilatation generator
and carry the conformal dimensions $\Delta_1,\Delta_2,\ldots, \Delta_n$. 
Three-point functions play a particular role since their form is fixed by
conformal invariance and since they contain the information about the
structure constants $C_{i\, j\, k}$ 
which appear in the theory's operator product 
expansion. For appropriately normalized conformal operators the three-point
functions take the form 
\begin{equation}
\langle {\cal O}_{\Delta_1}(x_1){\cal O}_{\Delta_2}(x_2)
{\cal O}_{\Delta_3}(x_3)\rangle= 
\frac{C_{\Delta_1\, \Delta_2\, \Delta_3}}{(x_{1}-x_2)^{\Delta-2\Delta_3}
(x_{2}-x_3)^{\Delta-2\Delta_1}(x_{3}-x_1)^{\Delta-2\Delta_2}}, 
\end{equation}
where $\Delta=\Delta_1+\Delta_2+\Delta_3$.

\subsection{Results on multi-point functions \label{results2}}

Before the advent of the BMN paper in 2002~\cite{Berenstein:2002jq}
results on multi-point functions mostly had to do with 
protected versions of these. A nice review
and a complete list of references can be found in~\cite{D'Hoker:2002aw}. 
  Here we will only very briefly list the
pre-BMN results. First, two- and 
three- point functions of 1/2 BPS and 1/4 BPS operators do 
not renormalize. Secondly, a large class of multi-point functions 
of 1/2 BPS operators have
very simple renormalization properties. These are the so-called 
extremal, next-to-extremal and near extremal correlators. Extremal 
correlators fulfill that $\Delta_1=\Delta_2+\ldots+\Delta_n$ and can 
always be expressed entirely in terms of two-point functions. 
Next-to-extremal correlators obey $\Delta_1=\Delta_2+\ldots +\Delta_{n} -2$
and factorize into a product of $n-3$ two-point functions and one
three-point function. Finally, near extremal multi-point functions have the
property that $\Delta_1=\Delta_2+\ldots +\Delta_{n}-2m$, where 
$2\leq m\leq n-3$ and $4\leq \Delta_1\leq 2n-2$. These multi-point 
functions can all be expressed in terms of lower point functions. 
The results on multi-point functions, briefly reviewed here, can also
be understood from the
string theory side~\cite{D'Hoker:2002aw}.

With the advent of the BMN limit~\cite{Berenstein:2002jq} the focus was shifted
from BPS operators to near BPS operators or BMN operators. 
As mentioned above these 
are operators which are created from long BPS operators by
the insertion of a few impurities. 
A much studied  set of  BMN operators belonging to the $SO(6)$ sector are the
following ones
\begin{equation} \label{BMN}
{\cal O}_{i\,j,n}^J= \frac{1}{\sqrt{J N^{J+2}}}
\left(\sum_{p=0}^n e^{\frac{2\pi i n}{J}}
\mbox{Tr}(\Phi_i\, Z^p \,\Phi_j \,Z^{J-p})-
\delta_{ij} \mbox{Tr}(\bar{Z}\, Z^{J+1})
\right),
\end{equation}
where $Z$ is one of the three complex scalars of ${\cal N}=4$ SYM, say
$Z=\Phi_1+i \Phi_2$, and  $i,j\in \{3,4,5,6\}$. These operators
are determined by 
the requirement that they should be
eigenvectors of the one-loop planar dilatation
operator~\cite{Berenstein:2002jq} in the limit $J\rightarrow \infty$.
(For the exact finite $J$ version of~(\ref{BMN}), see~\cite{Beisert:2002tn}.)
They can be organized into representations of $SO(6)$ in
the obvious way. The calculation of
 three-point functions of non-protected operators
such as BMN operators necessitates a highly non-trivial resolution of
operator mixing. First, in the case of extremal correlators, in order 
to calculate the classical three-point function to
leading order in $1/N$  one needs to take into account mixing between
single and double trace states~\cite{D'Hoker:1999ea}.
For BMN operators this calculation was carried out in
reference~\cite{Beisert:2002bb,Constable:2002vq} with the following result
for the space-time independent part of the three-point functions involving
two BMN operators and one 1/2 BPS operator of the form 
${\cal O}^J=\frac{1}{\sqrt{J N^J}}\mbox{Tr}( Z^J)$.
\begin{eqnarray}
\langle
\bar{{\cal O}}_{i j,n}^J \,
\,{\cal O}_{k l,m}^{r\cdot J} \, {\cal O}^{(1-r)\cdot J}
\rangle & =& \frac{2\, J^{3/2}\, \sqrt{1-r} \,\sin^2(\pi n r)}
{N \sqrt{r} \, \pi^2 (n^2-m^2/r^2)^2}
\left(1-\frac{\lambda(n^2-m^2/r^2)}{2J^2}\right) \times \nonumber \\
&& \hspace*{1.0cm}
\left(
\delta_{i(k}\delta_{l)j} n^2
+\delta_{i[k}\delta_{l]j} \frac{n m}{r}
+\frac{1}{4} \delta_{ij} \delta_{kl} \frac{m^2}{r^2}
\right),\label{threepoint}
\end{eqnarray}
where it is understood that the operators appearing
on the left hand side of~(\ref{threepoint})  have been redefined
to take into account the effects of the just mentioned
operator mixing.\footnote{Notice that in 
references~\cite{Kristjansen:2002bb,Constable:2002hw,Chu:2002pd} 
where classical
three-point functions of BMN operators also appear the contribution
to the three-point function from the mixing with double trace states was
{\it not} taken into account.}
To determine the order $\lambda$ correction  to the
structure constants requires a number of considerations. 
First, one actually has to resolve the operator mixing problem
to two loop order~\cite{Arutyunov:2002rs}, see also the discussion 
in~\cite{Alday:2005nd} as 
well as the remarks in~\cite{Beisert:2002bb,Constable:2002vq}.
The reason is that whereas the diagonalization of the dilatation generator
to one-loop order does not introduce any coupling constant dependent 
mixing of the states this is not so at two-loop order. At one-loop 
order one has a set of states $\{{\cal O}_{\alpha}\}$ 
which are simultaneously eigenstates at the classical and one-loop level.
However, when two-loop corrections are taken into account these eigenstates
are changed 
to $\{{\cal O}_{\alpha}+\lambda c_{\alpha \beta}{\cal O}_{\beta}\}$.
The coupling constant dependent modification of the states which occur
at two-loop level gives contributions to the structure constants
of order $\lambda$. Finally, one of course
has to ensure that the structure constants one reads off from the 
three-point functions are renormalization scheme independent. This
can be achieved by normalizing the two-point functions of the operators 
involved
to unity at order $\lambda$, see discussion in~\cite{Alday:2005nd}.


The early papers which dealt with three-point functions ignored either one 
or
both the two complications from operator
mixing, i.e.\ the mixing with multi-trace states and the mixing which 
naively appears to be of higher order.
References~\cite{Georgiou:2008vk,Georgiou:2009tp} dealt with 
the second type of mixing 
phenomenon and suggested to solve it using purely algebraic means, hence
avoiding the explicit evaluation of higher loop two-point functions.
References~\cite{Okuyama:2004bd,Roiban:2004va,Alday:2005nd} 
which studied one-loop
properties of structure constants did not take into account any of the
two above mentioned mixing issues. 
However, these references pointed out certain
connections of three-point functions to integrable spin chains
which we will review below together with some very recent progress along
the same lines~\cite{Escobedo:2010xs}.


\subsection{Multi-point functions and integrability}

As explained above calculating three-point functions involves first 
dealing with a subtle mixing problem and secondly executing the Wick
contractions between the appropriate eigenstates. We will follow the
historical development and postpone the discussion of the 
mixing problem to the end of
this section.
 
 For one-loop three-point functions of scalar
operators  one has tried to
derive a kind of effective vertex
which when applied to the three operators involved gives 
the order $\lambda$ contribution to the structure 
constant~\cite{Okuyama:2004bd,Alday:2005nd}. 
When evaluating 
three-point functions (apart from non-extremal ones) one generically
encounters two types of Feynman diagrams. One type
is two-point-like involving only non-trivial contractions 
between fields from two of the 
three operators appearing in the three-point function whereas 
the other
type involves non-trivial contractions between 
fields from all three operators. 
The generic term of the
effective vertex of~\cite{Alday:2005nd} correspondingly 
acts on the indices of three different operators.
However, 
one can show that
in a certain renormalization scheme the one-loop correction to
the structure constant only obtains contributions from Feynman
diagrams which are two-point-like~\cite{Okuyama:2004bd} and therefore
it is possible to construct an effective vertex whose terms 
act at most on indices from two different operators at a 
time~\cite{Okuyama:2004bd}. Both of the resulting effective
vertices have a close resemblance to the Hamiltonian
of the integrable $SO(6)$ spin chain. Notice, however, that 
both approaches~\cite{Okuyama:2004bd,Alday:2005nd} ignore
the two particular mixing issues discussed in the previous section.

An approach to the calculation of three-point functions which 
explicitly exploits the integrability of the planar dilatation generator
was presented in 
reference~\cite{Roiban:2004va}. Here the field theoretic 
three-point functions are represented as
matrix elements of  certain spin operators of the integrable 
spin chain determining the spectrum
and it is shown how these matrix elements can in 
principle be expressed in terms of  the elements of the 
spin chain's monodromy matrix. 
The method does not allow one to resolve the mixing between single and
multi-trace operators, however.

More recently, it was understood how, for
a certain subclass of operators, the mixing due to one-loop corrections 
and the calculation of tree-level three-point functions
could be efficiently 
dealt with using integrability tools having their origin in the 
planar integrability of the theory and this  led to exact results for
a  class of tree-level structure constants~\cite{Escobedo:2010xs}. Furthermore,
combining these tools with the ideas 
of~\cite{Roiban:2004va} a wealth of new data
on one-loop three-point functions for short operators was 
obtained~\cite{Escobedo:2010xs}. Notice again that these studies are
restricted to cases without mixing between single and multi-trace operators.
Reference~\cite{Grossardt:2010xq}
also contains extensive data on one-loop 
three-point functions for short operators
but here even the single trace mixing problem was not fully resolved
for all cases.

\subsection{Comparison to string theory}
Given the success of the comparison of the 
anomalous dimensions of gauge theory operators
with the energies of string states it is natural to look for a  
representation of the structure constants entering the
three-point functions
of non-protected operators in terms of string theory quantities.
With the discovery of the pp-wave limit of the type IIB string theory
and the corresponding BMN limit of ${\cal N}=4$ SYM hope was raised
that in this limit the AdS/CFT dictionary could be extended to include
the structure constants of the gauge theory and a first proposal for
the translation of these into string theory was put forward 
in~\cite{Constable:2002hw}. Here some structure constants $C_{ijk}$
were suggested to be related in a simple way to the matrix elements of the 
three-string vertex of the light cone string field theory. A lot of debate 
followed this initial proposal. First of all it was debated whether the 
$C_{ijk}$  were supposed to be the true CFT structure constants appearing
after taking into account the two types of operator mixing discussed in 
section~\ref{results2} or if the translation to string theory would not involve
this mixing. Secondly, as mentioned in section~\ref{comparison1} the exact
form of the three-string vertex of LCSFT was also a subject of debate.
The status of the discussion by the end of 2003 is well summarized in 
the review~\cite{Russo:2004kr}. In 2004 reference~\cite{Dobashi:2004nm}
provided a unifying description of the various earlier approaches. The
true LCSFT vertex was argued to be a linear combination of the two earlier
proposed ones and the $C_{ijk}$'s of relevance for the comparison between
gauge and string theory were argued to be the true CFT structure constants.
The precise translation of the gauge theory structure constants to the 
string theory language is well explained in~\cite{Dobashi:2004ka}.
All this should, however, be taken with some caution, as it has been 
understood that only for the full AdS/CFT system can one hope
for a complete matching of string and gauge theory, 
cf.\ the discussion in section~\ref{comparison1}.

In the past year there has been quite some progress in the calculation of 
two- and three-point correlation functions of string states 
in the full $AdS_5 \times S^5$ geometry using 
semi-classical methods. First, in~\cite{Janik:2010gc}
(see also~\cite{Dobashi:2002ar,Yoneya:2006td,Tsuji:2006zn}) a semi-classical
approach was shown 
to reproduce the characteristic conformal
scaling of the two-point function with the energy
for spinning strings with large quantum numbers 
and it was suggested that a similar approach could be applied to
three point functions. 
In~\cite{Buchbinder:2010vw} 
the semi-classical calculation of
two-point functions was formulated in terms of vertex operators
describing classical spinning strings~\cite{Tseytlin:2003ac,Buchbinder:2010gg}.
Subsequently, the semi-classical approach was extended to 
the calculation of three-point
functions involving two heavy states and one BPS 
state~\cite{Zarembo:2010rr,Costa:2010rz} and various cases of this type
were considered~\cite{Hernandez:2010tg,Ryang:2010bn,Georgiou:2010an}.
Furthermore, using the vertex operator representation of the correlation
functions a number of three-point functions between two heavy states 
and one light
non-BPS state was determined~\cite{Roiban:2010fe}. So far an explicit
comparison of the string theory three-point functions discussed here
and gauge theory three point functions has only been possible for 
protected correlators. However, very recently it has been suggested
that an expansion of the string theory three-point functions in a large
angular momentum of the heavy states might allow for a comparison with a 
gauge
theory perturbative expansion of the same quantity,
at least for the first few loop 
orders~\cite{Russo:2010bt}.

\section{Maldacena-Wilson loops \label{Wilsonloops}}

Wilson loops constitute an important
class of gauge invariant non-local observables in any gauge theory.
The idea that Wilson loops should have a dual string representation
has a long history, see~\cite{Polyakov:1997tj} and references therein.
A realization of this idea in the context of the
AdS/CFT correspondence was obtained by Maldacena who
introduced the following special type of locally supersymmetric Wilson loops
~\cite{Maldacena:1998im}
\begin{equation}
W[C]=\frac{1}{\mbox{dim}({\cal R})}\mbox{Tr}_{\cal R}\,
\left({\mbox P}\,
\exp \left[ \oint_C d\tau \left(
iA_{\mu}(x)\dot{x}^{\mu}+\Phi_i(x) \theta^i|\dot{x}|
\right)\right]\right).
\end{equation}
Here ${\cal R}$ denotes an irreducible representation of $SU(N)$, 
$x^{\mu}(\tau)$ is a parametrization of the loop $C$, $\Phi_i(x)$ are
the 6 real scalar fields of ${\cal N}=4$ SYM and $\theta_i(\tau)$ is a 
curve on 
$S^5$. In the present section we will use the following definition of
the 't Hooft coupling constant
\begin{equation}
\lambda=g_{{\mbox{\tiny YM}}}^2 \,N.
\end{equation}
According to Maldacena~\cite{Maldacena:1998im}
the expectation value of such a Wilson loop in the fundamental
representation should be
determined by the action of
a string ending at the curve $C$ at the boundary of $AdS_5$, i.e.
\begin{equation}
\langle W[C]\rangle
=\int_{\partial X=C} {\cal D}X \exp \left(-\sqrt{\lambda}S[X]\right).
\label{stringrecipe}
\end{equation}
Expectation values of many supersymmetric 
Wilson loops have turned 
out to be expressible in terms of expectation values in 
integrable zero-di\-men\-sional matrix models.
Furthermore, Wilson loops
have provided us with the most promising test of the
AdS/CFT correspondence beyond the planar limit to date. The relation
between Maldacena-Wilson loops and spin chain integrability is
so far rather sparse, cf.\ subsection~\ref{instances}.

\subsection{The 1/2 BPS line and circle \label{BPS}}

A Wilson loop in form of a single straight line, i.e.given by \ $x(\tau)=\tau,$ 
$\theta^i(\tau)=const$,
constitutes a 1/2 BPS object. 
Its expectation value does not get any
quantum corrections and is exactly equal to one. The circular Wilson
loop parametrized by
\begin{equation}
x(\tau)=(\cos \tau,\sin \tau, 0,0),
\end{equation}
and $\theta^i(\tau)= const$ can be obtained from the straight line by
a conformal transformation and is likewise 1/2 BPS.
Its expectation value does get quantum corrections, however.
The expectation value of the circular
Wilson loop was calculated at the planar level in perturbation
theory to
two loop order
 in~\cite{Erickson:2000af} and it was found that only ladder like 
diagrams
(i.e.\ diagrams whose vertices all lie on the loop) contributed. 
The authors of~\cite{Erickson:2000af} proposed that 
this could be true to all orders and showed that
under that assumption the calculation of the expectation value could be
reduced to a combinatorial problem the answer to which was given by
an expectation value in a zero-dimensional Gaussian matrix model. 
Subsequently, it was understood that the reason why the problem was
zero-dimensional in nature was that the expectation value of the 
circular
Wilson loop could be understood as an anomaly arising at the point
at infinity when 
conformally mapping the straight line to a circle~\cite{Drukker:2000rr}.
In addition, the proposal of~\cite{Erickson:2000af}
was extended to all orders in the $\frac{1}{N}$-expansion
~\cite{Drukker:2000rr}. Stated precisely, the proposal says that the 
expectation value of the circular Wilson loop is given to all orders
in $\lambda$ and all orders in $\frac{1}{N}$ by the following 
expression~\footnote{Here the integration is over Hermitian matrices, i.e.\
${\cal D}M= \prod_i dM_{ii} \prod_{j>i} d \Re(M_{ij}) d \Im(M_{ij})$ and
$Z$ is the partition function of the model.}
\begin{equation}
\langle W_{circle}\rangle=
\langle \frac{1}{N} \mbox{Tr}\left(\exp(M)\right)\rangle
= \frac{1}{Z} \int {\cal D} M \frac{1}{N} \mbox{Tr} \left(\exp (M)\right) 
\exp\left(-\frac{2N}{\lambda} \mbox{Tr}\, M^2\right).
\label{circular}
\end{equation}
Using matrix model techniques the expectation value can be calculated
{\it exactly} and yields~\cite{Drukker:2000rr}
\begin{equation}
\langle W_{circle}\rangle = \frac{1}{N}\,L_{N-1}^1 (-\lambda/4N)
\exp(\lambda/8N), \label{circle}
\end{equation}
where $L_{N-1}^1$ is a Laguerre polynomial. One can explicitly write down
the genus expansion of~(\ref{circle})
and then taking the strong coupling, $\lambda\rightarrow \infty$, limit of this
one gets
\begin{equation}
\langle W_{circle}\rangle =\sum_{p=0}^{\infty} \frac{1}{N^{2p}}\,
\frac{e^{\sqrt{\lambda}}}{p!}\sqrt{\frac{2}{\pi}}\,
\frac{\lambda^{\frac{6p-3}{4}}}{96^p}\,
\left[
1-\frac{3(12p^2+8p+5)}{40\sqrt{\lambda}}+{\cal O}
\left(\frac{1}{\lambda}\right)
\right].\label{gauge}
\end{equation}
The
possibility of the expectation value getting additional contributions
from instantons was investigated in~\cite{Bianchi:2001jg,Bianchi:2002gz}.
Recently, however, the proposal of~\cite{Erickson:2000af,Drukker:2000rr} was 
proved to be true~\cite{Pestun:2007rz}. 

The expectation value of the circular Wilson loop can be found from the string
theory recipe~(\ref{stringrecipe}) in the strong coupling limit by performing
a saddle point analysis. It turns out that the string action is dominated by
its bosonic part at the saddle point and the calculation becomes equivalent
to determining the area of the minimal area surface ending at the loop $C$. 
The minimal surface area, however, 
diverges and requires a regularization which 
results in the saddle point action being negative~\cite{Maldacena:1998im}. 
The minimal area 
corresponding to the circle was first determined 
in~\cite{Berenstein:1998ij,Drukker:1999zq} and led to the first crude
estimate of the expectation value of the planar circular Wilson loop from the
string theory side $\langle W_{circle}\rangle^{string}\sim e^{\sqrt{\lambda}}$.
Later the string analysis was extended to include sub-leading corrections
in $\lambda$ coming from integration over zero-modes and to include higher
genus surfaces~\cite{Drukker:2000rr}. This led to the following 
string theory estimate
of the expectation of the circular Wilson loop 
\begin{equation}
\langle W_{circle}\rangle^{string} 
\propto\sum_{p=0}^\infty \frac{1}{N^{2p}}\, 
\frac{e^{\sqrt{\lambda}}}{p!}\,\,\lambda^{\frac{6p-3}{4}}\,
\left[1+{\cal O}\left(\frac{1}{\sqrt{\lambda}}\right)
 \right]. \label{string}
\end{equation}
The matching between~(\ref{gauge}) and~(\ref{string}) provides a piece of 
evidence in favour of the validity of
the AdS/CFT correspondence 
beyond the planar level. In order to reproduce the additional factor
$\sqrt{\frac{2}{\pi}}$ appearing 
in~(\ref{gauge}) from string theory one needs to take into account 
the  
fluctuations about the minimal surface. The framework for performing this
calculation at the planar level
was laid out in~\cite{Drukker:2000ep} and recently interesting 
progress was achieved in the explicit evaluation of the missing
sub-leading  
contribution in the planar case \cite{Kruczenski:2008zk}. 

\subsection{More supersymmetric Wilson loops}

In reference~\cite{Zarembo:2002an} Zarembo found a series of 
Wilson loops of $1/4$, $1/8$ and $1/16$ BPS type which
can be viewed as generalizations of the 1/2 BPS Wilson line living in
the higher dimensional subspaces $\mathrm{I}\!\mathrm{R}^2$, 
$\mathrm{I}\!\mathrm{R}^3$ and $\mathrm{I}\!\mathrm{R}^4$.
These Wilson loops all 
have trivial expectation values. This was argued from the gauge
theory side in~\cite{Zarembo:2002an,Guralnik:2003di} and an understanding
from the string theory perspective was provided in~\cite{Dymarsky:2006ve}.
Finally, it was explained by topological arguments in~\cite{Kapustin:2006pk}.

The first example of a 1/4 BPS Wilson loop with non-trivial expectation value
was found by Drukker~\cite{Drukker:2006ga}. Later a large family of
supersymmetric Wilson loops with non-trivial expectation values was
identified~\cite{Drukker:2007dw,Drukker:2007yx,Drukker:2007qr}. This family
of loops constitute generalizations of the 1/2 BPS circular loop above. 
The most generic type is 1/16 BPS and lives on an $S^3$ sub-manifold
of four-dimensional space-time. Loops further restricted
to an $S^2$ are 1/8 BPS and their expectation values were conjectured
 to be equal
to the analogous expectation values in the zero instanton
sector of two-dimensional Yang-Mills
theory on a sphere~\cite{Drukker:2007yx,Drukker:2007qr} which implies
that they can again be evaluated using a matrix model.
 More precisely,
for such loops we should have
\begin{equation}
\langle W[C]\rangle =
\frac{1}{N} L_{N-1}^1\left(g_{{\mbox{\tiny YM}}}^2
 \frac{{\cal A}_1 {\cal A}_2}{{\cal A}^2}
\right)
\exp\left[-\frac{g_{{\mbox{\tiny YM}}}^2}{2}\frac{{\cal A}_1 {\cal A}_2}{{\cal A}^2}\right],
\end{equation}
where ${\cal A}_1$ and ${\cal A}_2$ are the two areas of the sphere bounded
by the loop and ${\cal A}={\cal A}_1+{\cal A}_2= 4\pi$. 
Perturbative gauge theory a
arguments supporting the conjecture were presented 
in~\cite{Drukker:2007yx,Drukker:2007qr,Bassetto:2008yf,Young:2008ed} and
string theoretic arguments in favour of the conjecture
appeared in~\cite{Giombi:2009ms}. The conjecture was further supported
by studies using localization techniques in~\cite{Pestun:2009nn}. 

A unifying and 
exhaustive description of all supersymmetric Wilson loops was given 
in~\cite{Dymarsky:2009si} and it was found that the two classes of 
Wilson loops described by respective Zarembo and Drukker et al.\ are indeed
the two most natural ones.

Some aspects of the analysis
outlined above have been generalized to ${\cal N}=6$ supersymmetric
Chern-Simons matter theory. The 1/2 BPS Wilson loop has been 
constructed~\cite{Drukker:2009hy}
and
its expectation value shown to be expressible in terms of an expectation
value in a zero-dimensional
supermatrix model~\cite{Kapustin:2009kz,Drukker:2009hy}.
In addition, one has identified a 1/6 BPS Wilson 
loop~\cite{Drukker:2008zx,Chen:2008bp,Rey:2008bh} whose expectation value
can likewise be calculated using a matrix 
model~\cite{Kapustin:2009kz,Marino:2009jd}. 

\subsection{Higher representations}

Having obtained the result~(\ref{circular}) and using the Schur polynomial
formula
one has access to the expectation
value 
of the 1/2 BPS circular Wilson
loop in any given irreducible representation of $SU(N)$. When the 
rank of the representation, $k$, i.e.\ the number of boxes
in the Young tableau, fulfills that $k\sim {\cal O}(N)$ the appropriate
string theory description of the Wilson loop is in terms of Dp-branes 
rather than fundamental
strings.
 Early ideas in this direction were presented 
in~\cite{Callan:1997kz,Rey:1998ik,Drukker:2005kx}.
 The precise dictionary between Wilson
loops in higher representations and Dp-branes was 
found in~\cite{Gomis:2006sb}. A Wilson loop operator in a  
representation given by a Young diagram with $M$ rows and $K$ columns
with $n_i$ boxes in the $i$'th row and $m_j$ boxes in the $j$'th column
has two different string realizations. One is in terms of $K$ D3-branes
carrying electric charges $n_1,\ldots  ,n_K$ and the other is in terms 
of M D5-branes carrying electric charges $m_1,\ldots , m_M$. In both
cases, as long as $k\ll N^2$, one should be able to 
determine the expectation value of the
Wilson loop by treating the Dp-brane using the probe approximation,
i.e.\ ignoring the back reaction of the $AdS_5\times S^5$ 
geometry.~\footnote{ 
In particular, it is expected that
energies of certain spinning D3- and D5-branes  correspond to anomalous 
dimensions of local twist operators (cf.\ the chapter~\cite{chapTwist})  
carrying higher 
representations of the gauge group~\cite{Armoni:2006ux,Gang:2009pk}.}

For the completely symmetric and the completely antisymmetric
representation of rank $k$ the gauge theory expectation value of
the 1/2 BPS circular Wilson loop has been extracted from the matrix model in 
the limit $N\rightarrow \infty$, $k\rightarrow \infty$
with $k/N$ fixed using saddle point techniques. In the antisymmetric case
the result in the large $\lambda$ limit reads~\cite{Hartnoll:2006is}
\begin{equation}
\langle W_{A_k}(C)\rangle =\exp\left[\frac{2N}{3\pi} \sqrt{\lambda} \sin^3 \theta_k\right],
\end{equation}
where $\theta_k$ is given by
\begin{equation}
\pi\left(1-\frac{k}{N}\right)=\left(\theta_k-\sin(\theta_k) \cos(\theta_k)
\right).
\end{equation}
This result matches the result of a supergravity calculation on the 
string theory side using D5-brane probes~\cite{Yamaguchi:2006tq}.
For the completely symmetric representation the situation is more
involved since in the large $N$ analysis
one encounters two different saddle points. Which one
dominates depends on the on the precise values of $\lambda$ and $k/N$.
If one considers the limit of large $\lambda$ and $N$ with a fixed value
of $\kappa$, defined by
\begin{equation}
\kappa=\frac{\sqrt{\lambda}k}{4N},
\end{equation}
one finds~\cite{Hartnoll:2006is,Okuyama:2006jc}
\begin{equation}
\langle W_{S_k}^{(1)}[C]\rangle=\exp\left[2N \left(\kappa \sqrt{1+\kappa^2}+\sinh^{-1}(\kappa)
\right)
\right].
\end{equation}
This result matches a supergravity calculation carried out using 
D3-brane probes\cite{Drukker:2005kx}. The same saddle point dominates
in the limit $\lambda\rightarrow\infty$, $k\rightarrow \infty$,
$N\rightarrow \infty$ with $k/N$ fixed. In other regions of the 
parameter space the second saddle point might come into play and in 
general one has that the expectation value of the Wilson loop
in the symmetric representation is a sum of two terms, i.e.
$W_{S_k}[C]=W_{S_k}^{(1)}[C]+W_{S_k}^{(2)}[C]$.

When the rank of the representation reaches the size $k\sim {\cal O}(N^2)$ 
the probe
approximation breaks down as the back reaction of the $AdS_5\times S^5$
geometry can no longer be ignored. In this case the resulting string background
can be described as a bubbling geometry~\cite{Lin:2004nb}. The 
determination of
the bubbling geometry corresponding to  1/2 BPS Wilson loops was initiated  
in~\cite{Yamaguchi:2006te,Lunin:2006xr} and completed
in~\cite{D'Hoker:2007fq}. The calculation of the expectation value of
the Wilson loop from the gauge theory side
still proceeds via the matrix model and was 
carried out in~\cite{Okuda:2007kh,Okuda:2008px}.

Like the
1/2 BPS Wilson loop the less supersymmetric Wilson loops
can be studied in higher representations of the gauge group. 
This was done for a number of 1/4 BPS Wilson loops in~\cite{Drukker:2006zk}.
There also exist numerous results on correlation functions involving multiple
Wilson loops as well as Wilson loops and local operators for loops
in various representations.

\subsection{Other instances of integrability of Wilson loops
\label{instances}}

As explained in section~\ref{BPS} expectation values of Wilson loops 
in the strong coupling, $\lambda\rightarrow \infty$ limit can be evaluated
by finding a classical string solution with appropriate boundary conditions. 
The string sigma model describing type IIB strings on $AdS_5\times S^5$
is known to be classically integrable~\cite{Mandal:2002fs,Bena:2003wd} 
and this fact was exploited in reference~\cite{Drukker:2005cu}
to find the strong coupling 
expectation values of numerous Wilson loops with $x^\mu(t)$ and $\theta^i(t)$
periodic. More recently a class of polygonal 
(non-supersymmtric) Wilson loops built from light
like segments have attracted attention due to their relation with gluon
scattering amplitudes~\cite{Alday:2007hr}. 
The minimal surfaces corresponding to these
loops have turned out to be described by integrable systems of Hitchin
type. For a discussion of Wilson loops related to scattering amplitudes
and the relevant set of references we
refer to the chapters~\cite{chapAmp,chapDual,chapTDual}.

It seems difficult to relate the expectation  value
of supersymmetric Wilson loops to integrable
spin  chains but there exists one special
construction which exposes such a relation.
In reference~\cite{Drukker:2006xg} the authors studied insertion of
composite operators into Wilson loops. The Wilson loop was taken
to be a straight line or a circle and  $\theta^i$ to describe a single point
on $S^5$. Furthermore, the
composite operator was assumed to be built from two complex scalars
$Z=\left( \Phi_1+ i \Phi_2\right)/\sqrt{2}$ and
$X=\left(\Phi_3+i\Phi_4\right)/\sqrt{2}$. It is possible to assign a 
conformal dimension to such an inserted operator and to determine this
dimension one has to solve a certain mixing problem involving two-point
functions of the type
\begin{equation}
\langle W_{line}\left[{\cal O}_{\beta}^{\dagger}(t){\cal O}_{\alpha}(0)
\right] \rangle
=\langle \frac{1}{N}\mbox{Tr} 
\left( P \, {\cal O}_{\alpha}^{\dagger}(t){\cal O}_{\beta}(0)
\exp\left[i\int(A_t+i\, \Phi_6)dt \right] \right )
\rangle.
\end{equation}
An operator insertion ${\cal O}_{\Delta}$ with a well-defined conformal
dimension fulfills
\begin{equation}
\langle W_{line}\left[{\cal O}_{\Delta}^{\dagger}(t){\cal O}_{\Delta}(0)
\right] \rangle \sim \frac{1}{t^{2\Delta}}.
\end{equation}
The above mixing problem was studied at the planar one-loop order 
in~\cite{Drukker:2006xg} and  mapped onto
the problem of diagonalising the Hamiltonian of
an $SU(2)$ open Heisenberg spin chain
with completely reflective boundary conditions. This spin chain is integrable
and can be solved by Bethe ansatz. For a description of the Bethe equations
associated with integrable open spin chains, we refer to 
the chapter~\cite{chapDeform}.
The string dual of the inserted operator
can be identified and a successful comparison between the gauge theory
side and string theory side for inserted operators of BMN type and of 
the type dual to spinning
strings was carried out in~\cite{Drukker:2006xg}.

\section{Conclusion}

The search for spin chain like integrable structures in ${\cal N}=4$ SYM 
regarding non-planar anomalous dimensions
and Maldacena-Wilson loops has so far not 
provided us with very strong positive results.
Maldacena-Wilson loops are more naturally related to zero-dimensional 
integrable matrix models than to spin chains and non-planar anomalous
dimensions have not yet provided us with any traces of integrability.
 It is possible
that one can learn more about non-planar anomalous dimensions by studying
the three-point functions or structure constants 
of the theory. Non-trivial operator mixing issues, however, make the evaluation
of structure constants quite involved. For a subset of single trace
operators the mixing
is an entirely planar effect and can in principle
be handled using tools originating from
the planar integrability of the theory. In the generic case, however, 
single trace operators will mix with multi-trace operators and the 
calculation of structure constants requires a diagonalization of
the non-planar dilatation operator.
The most naive approach to studying non-planar anomalous 
dimensions, namely doing perturbation theory in $\frac{1}{N}$ requires 
dealing with the splitting and joining of spin-chains and leads to a Hilbert
space of states for which the standard concepts of integrability such
as the asymptotic S-matrix and two-particle scattering do not 
immediately apply. Going beyond the planar limit 
hence seems to require a
rethinking of the entire framework of integrability or invoking
some non-perturbative way of handling the higher topologies.

\vspace*{0.7cm}

\noindent
{\bf Acknowledgments:}
I am grateful to N.\ Beisert, N.\ Drukker, 
C.\ Sieg, A.\ Tseytlin, D.\ Young and K.\ Zoubos 
for useful discussions and comments during the writing of this review.
This work was supported by FNU through grant number 272-08-0329.

\phantomsection
\addcontentsline{toc}{section}{\refname}
\bibliography{chapIV1,chapters}
\bibliographystyle{nb}

\end{document}